\documentclass[12pt,a4paper,english]{revtex4}
\usepackage[T1]{fontenc}
\usepackage[latin1]{inputenc}
\usepackage{subfigure}
\usepackage{color}
\usepackage{graphicx}

\makeatletter

\providecommand{\LyX}{L\kern-.1667em\lower.25em\hbox{Y}\kern-.125emX\@}
\newcommand{\noun}[1]{\textsc{#1}}

\usepackage{babel}
\makeatother
\begin{document}

\title{Strain, size and composition of InAs Quantum Sticks, \emph{embedded}
in InP, by means of Grazing Incidence X-ray Anomalous Diffraction}

\author{A. Letoublon$^{1}$, V. Favre-Nicolin$^{1,2}$, H. Renevier$^{1,2}$\\
M.G. Proietti$^{3}$, C. Monat$^{4}$, M. Gendry$^{4}$\\
O. Marty$^{5}$, C. Priester$^{6}$}

\address{$^{1}$Commissariat à l'Energie Atomique, Département de Recherche
Fondamentale sur la Matière Condensée, SP2M/NRS, 17 rue des martyrs,
38054 Grenoble Cedex 9, France.}

\address{$^{2}$Université Joseph Fourier, BP 53, F-38041, Grenoble Cedex
9, France.}

\address{$^{3}$Departamento de Física de la Materia Condensada, Instituto
de Ciencia de Materiales de Aragón, CSIC-Universidad de Zaragoza -
c. Pedro Cerbuna 12, 50009 Zaragoza, Spain.}

\address{$^{4}$LEOM, UMR-CNRS 5512, Ecole Centrale de Lyon, 69134 Ecully,
France.}

\address{$^{5}$LENAC, Université Lyon I, 69621 Villeurbanne, France. }

\address{$^{6}$Institut d'Electronique, de Microélectronique et de Nanotechnologie,
dep. ISEN, 59652 Villeneuve d'Ascq, France.}

\email{Hubert.Renevier@cea.fr}

\date{22/09/03, submitted}

\begin{abstract}
We have used x-ray anomalous diffraction to extract the x-ray structure
factor of InAs quantum stick-like islands, \emph{embedded} in InP.
The average height of the quantum sticks (QSs), as deduced from the
width of the structure factor profile is 2.54nm. The InAs out of plane
deformation, relative to InP, is equal to 6.1\%. Diffraction Anomalous
Fine Structure provides a clear evidence of pure InAs QSs. Finite
Difference Method calculations reproduce well the diffraction data,
and give the strain along the growth direction. Chemical mixing at
interfaces is at most of 1ML

\end{abstract}
\maketitle
Band structure gives the optical and electronic properties of materials.
It can be strongly modified by reducing the size of semiconductor
material \textcolor{red}{}down to a length scale comparable to the
effective wave-length of the carriers\textcolor{black}{, i.e. in the
order of several} nanometers, leading to discrete energy levels \cite{Bimberg99}.
\textcolor{black}{Much study has been done on InAs quantum structures
such as Quantum Wires (QWrs) and Quantum Dots (QD). With an emission
wavelength which can be around 1.55$\mu $m, such nanostructures are
of interest for the next generation of integrated circuits (telecommunication
relevant range of 1.3-1.6$\mu $m).} The nano-objects we are interested
in are grown by Molecular Beam Epitaxy (MBE) and obtained via the
Stranski-Krastanov growth. The lattice mismatch $\frac{a_{InAs}-a_{InP}}{a_{InP}}$
is about 3.2\%. Recent studies \cite{Gonzalez00} have shown that
a strong stress anisotropy appears during the growth deposition leading
to a higher stress along the {[}110{]} direction than along the $[1\bar{1}0]$.
The stress is released first in the {[}110{]} direction leading to
the QWrs formation. To be suitable for devices, the nanostructures
are encapsulated with InP or embedded in a superlattice, they must
be homogeneous in size, shape and composition, to provide well defined
emission wavelengths. The knowledge of strain field, chemical gradients,
chemical mixing at the interface, is of great importance to understand
the growth dynamics as well as the electronic and optical properties
of the nanostructures.

\begin{figure}
\subfigure[]{\includegraphics[  width=0.45\textwidth,
  keepaspectratio]{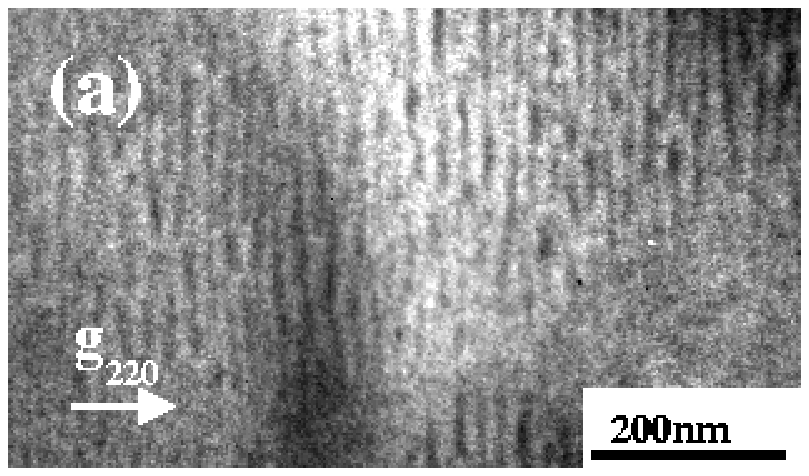}}\subfigure[]{\includegraphics[  width=0.45\textwidth,
  keepaspectratio]{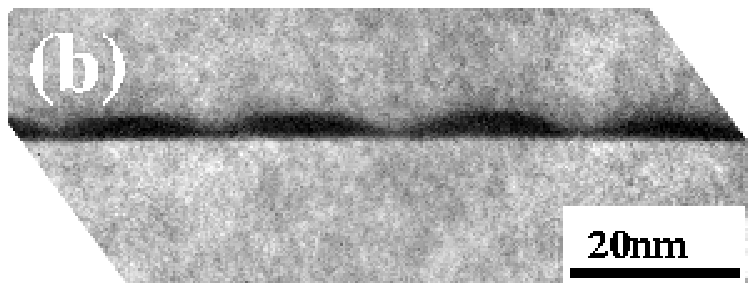}}

\caption{\label{cap:TEM pictures} Transmission Electron Microscopy images
of InAs QSs embedded in InP (a) plane view, (b) cross section ; one
can clearly see the truncated triangle shape}
\end{figure}

We report on an x-ray study of InAs stick-like islands embedded in
InP with 10 nm thick capping layer. These samples are obtained by
optimisation of the MBE growth parameters to minimize the As/P exchange
and reduce the height dispersion of the InAs islands \cite{Monat02,Gendry03}.
The width of photoluminescence (PL) peaks reveals a very good height
homogeneity \cite{Salem02}, as confirmed by Transmission Electron
Microscopy (TEM) measurements. The InAs islands obtained with a nominal
deposition thickness of 4ML are 50 to 200nm long along the $[1\bar{1}0]$
direction (\ref{cap:TEM pictures}a). As shown by cross-section TEM
image the QSs exhibit a truncated triangle side shape with typical
width of 22.5 \ensuremath{±} 0.12 nm and homogeneous height, of 2.4
nm (figure \ref{cap:TEM pictures}b).

In the present paper we show that the structure factor of embedded
nanostructures can be directly determined by means of grazing incidence
anomalous x-ray diffraction, allowing to recover the average height
and strain of the QSs, to determine their composition and check the
As/P exchange. As a matter of fact, strain of buried nanostructures
is not directly related to composition (Vegards's law), but it also
depends on size, morphology, and cap layer thickness. Then, tuning
the x-ray energy near an absorption edge of atoms that belong to the
nanostructures is a way to modify their scattering power and to enhance
the chemical sensitivity of diffraction. Previous studies have shown
the interest of using anomalous diffraction to determine strain and
composition in InAs QWrs grown on InP \cite{Grenier02} or large Ge
dots grown on Si \cite{Magalhaes02,Schulli03}. We report on a general
method that takes advantage of the full capability of anomalous diffraction
\cite{Hodeau2001} and can be applied to the very interesting and
challenging case of \emph{small size} \emph{embedded} nanostructures,
the x-ray scattering yield of which is overwhelmed by the dominant
matrix contribution, whatever the momentum transfer is.

Grazing incidence anomalous diffraction intensity, can be written
in the frame of the Distorted Wave Born Approximation \cite{Dosch92,Proietti99}
:

\begin{equation}
I(\vec{Q},E)\propto \left\Vert T_{i}(\alpha _{i},E)\right\Vert ^{2}\left\Vert F_{T}(\vec{Q})\right\Vert ^{2}\left[\left(\cos (\varphi _{T}-\varphi _{A})+\beta f_{As}^{\prime }\right)^{2}+\left(\sin (\varphi _{T}-\varphi _{A})+\beta f_{As}^{\prime \prime }\right)^{2}\right]\label{eq:I(Q,E)}\end{equation}
where $F_{T}$ is a complex structure factor of phase $\varphi _{T}$
which includes the overall contribution of non anomalous atoms and
the Thomson scattering of all anomalous atoms, $\beta =\frac{\left\Vert F_{A}\right\Vert }{f_{As}^{0}\left\Vert F_{T}\right\Vert }$
where $F_{A}$ is a complex structure factor of phase $\varphi _{A}$
which includes the Thomson scattering of all anomalous atoms (i.e.
As atoms). $\left\Vert T_{i}(\alpha _{i},E)\right\Vert ^{2}$ is the
incidence transmittivity, and $\alpha _{i}$ is the incidence angle
that is close to the critical angle $\alpha _{c}$. In the following,
the exit angle is much larger than the critical angle, then the exit
transmittivity $\left\Vert T_{f}(\alpha _{f},E)\right\Vert ^{2}=1$
and the scattering length ($\approx 100nm$) is much larger than the
epilayer thickness. The As scattering factor writes $f_{As}=f_{As}^{0}+f_{As}^{\prime }+if_{As}^{\prime \prime }$,
where $f_{As}^{\prime }$ and $f_{As}^{\prime \prime }$ are the real
and imaginary resonant scattering of As atoms. Equation \ref{eq:I(Q,E)}
shows that diffraction measurements at various energies (at least
3) at the As K-edge, allow to recover a quantity that is proportional
to the $F_{T}$ modulus, the $\beta $ ratio and the phase difference
$\Delta \varphi =\varphi _{T}-\varphi _{A}$. The knowledge of $\left\Vert F_{A}\right\Vert $and
$\Delta \varphi $ readily gives information about the average size
and strain of the nanostructures and chemical mixing at interface. 

\begin{figure}
\includegraphics[  width=0.55\textwidth,
  keepaspectratio]{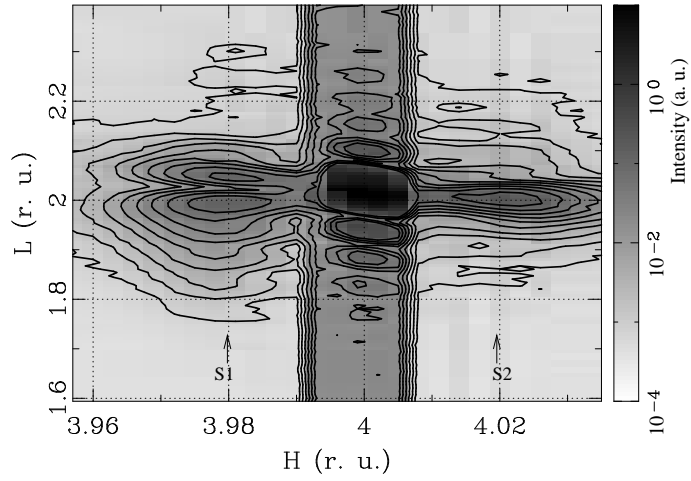}

\caption{\label{cap:Diffraction-map} Experimental diffraction map around
the weak (442) reflection recorded at 11.840 KeV in grazing incidence
geometry ($\alpha _{i}=\alpha _{c}=0.2°$, $\alpha _{f}\simeq 20°$).
S1 and S2 are correlation satellites due to the stick short range
periodicity.}
\end{figure}

Grazing incidence Anomalous diffraction at the As K-edge (11.867 KeV)
was performed at the French Collaborative Research Group beamlines
BM32 and BM2 at the European Synchrotron Radiation Facility, using
a Si(111) double crystal monochromator to select energies and mirrors
to reject harmonics. Grazing incidence was used to minimize the substrate
contribution. We recorded the scattering intensity in the vicinity
of the (442) substrate reflection, at several energies across the
As K-edge. We chose a weak reflection (h+k+\emph{l}=4n+2) for which
the anomalous diffraction contrast is maximum; it lies in the mirror
plane defined by the {[}110{]} and {[}001{]} directions to benefit
from symmetry in the reciprocal space. Figure \ref{cap:Diffraction-map}
shows the diffraction intensity map recorded at 11.840 keV and at
a grazing incidence angle equal to the critical angle ($\alpha _{i}=\alpha _{c}=0.2°$)
to maximize the diffraction intensity. For that reflection the exit
angle $\alpha _{f}$ is about 20°. A large amount of information can
be drawn from such a map : the spreading of scattering in the {[}110{]}
direction is due to both short range correlation and lattice strain
in the corresponding direction in the real space ({[}110{]}) whereas,
in the {[}001{]} direction it is due to the sharp strain evolution.
Figure \ref{cap:Diffraction-map} shows correlation satellites on
both sides of the (442) (S1 and S2 ) their positions relative to the
substrate peak give a qualitative estimation of the mean correlation
distance between the wires. We find a value of $20.7nm$, that is
in agreement with the TEM. 

The markable result in this map, is the clear asymmetry of the S1
and S2 satellites ; S1 exhibits a broad and twin feature along the
{[}001{]} direction, with a sharp splitting at \textbf{\emph{l}}=2,
whereas S2 is a single feature centered approximately at \textbf{\emph{l}}=2.
S1 corresponds to regions stretched along the {[}110{]} direction
(tensile strain), i.e. to InAs sticks and InP regions located below
and above the sticks. S2, instead, corresponds to InP which is compressed
in between of the sticks. Indeed, anomalous diffraction measurements
at the As K-edge show no significant intensity variations of satellite
S2 as a function of the energy, confirming that the InP contribution
is the dominant one. Furthermore, these measurements also show firstly,
that the transmittivity corrections can be neglected, secondly, that
the wetting layer, if it exists, must be very thin, at most one ML.
Cutoff fringes are clearly observed in the \textbf{\emph{l}} direction
around the substrate Bragg peak position and give the total thickness
of both the InAs QSs and the InP capping, that is equal to $11\pm 0.3nm$.

\begin{figure}
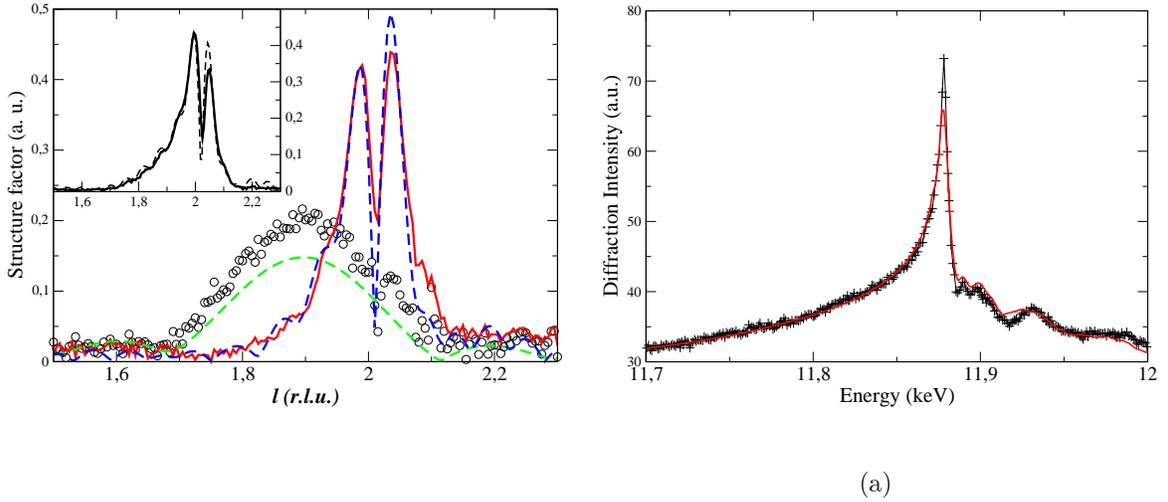

\includegraphics[  width=0.45\textwidth]{fig3a.eps}~~~~\subfigure[]{\includegraphics[  width=0.45\textwidth]{fig3b.eps}}

\caption{\label{cap:FT_FA} (a) Experimental $F_{T}$ (solid line) and $F_{A}$
(ooo) modulus as a function of \textbf{\emph{}}the reciprocal lattice
unit \textbf{\emph{l}} at \textbf{\emph{h}}=\textbf{\emph{k}}=3.98
(across the satellite S1). $F_{A}$ is the structure factor that corresponds
to all anomalous atoms, i.e. As atoms only. Also shown are best simulation
curves (dashed line) obtained with a FDM model made of pure InAs QSs
with a truncated triangle side profile. Inset shows the experimental
(solid line) and simulated (dashed line) square root of the diffraction
intensity at 11.867keV ; at this energy, anomalous diffraction is
maximized. (b) GI- DAFS spectrum recorded, at the As K-edge, at the
maximum of the $F_{A}$ profile (\textbf{\emph{l}}=1.9) and the best
fit curve obtained with pure InAs structure (solid line).}
\end{figure}

We performed \textbf{\emph{l}}-scans (\textbf{\emph{h}}=\textbf{\emph{k}}=3.98)
accross the satellite S1, at nine different energies close to the
As K-edge including those corresponding to the minimum of $f_{As}^{\prime }$
and the white line of $f_{As}^{\prime \prime }$. A weak fluorescence
background was substracted and the data were normalized to \textbf{\emph{l}}-scans
across the satellite S2, measured at the same energies. Then $F_{A}(\vec{Q})$,
$F_{T}(\vec{Q})$ and $\Delta \varphi (\vec{Q})$ were extracted by
fitting equation \ref{eq:I(Q,E)} to the energy dependent experimental
intensities.

Figure \ref{cap:FT_FA}a) shows the experimental modulus of $F_{A}(\vec{Q})$,
$F_{T}(\vec{Q})$. The QSs height average value ($\left\langle H\right\rangle $)
can be estimated from the Full Width at Half Maximum (FWHM) of $F_{A}$
, $\left\langle H\right\rangle =\frac{c_{InP}}{0.9\times (\Delta F_{A})_{FWHM}}=2.54nm$,
that is very near of the value measured with TEM. In order to determine
the QSrs composition and the local strain accomodation, Grazing Incidence
Diffraction Anomalous Fine Structure spectrum was measured at the
maximum of $F_{A}$ \textbf{\emph{(h}}=\textbf{\emph{k}}=3.98 and
\textbf{\emph{l}}=1.9) at the As K-edge. Figure \ref{cap:FT_FA}b)
shows the experimental DAFS spectrum and the simulation calculated
with the crystallographic structure of InAs$_{1-x}$P$_{x}$ and the
anomalous scattering factors $f_{As}^{\prime }$ and $f_{As}^{\prime \prime }$of
bulk InAs. A scale factor, the detector efficiency as a function of
the energy and the As occupation factor ($1-x$) were refined. The
best fit curve, shown on figure \ref{cap:FT_FA}b) corresponds to
a value of ($1-x$) equal to 1.03, i.e. the QSs composition is \emph{pure}
InAs. GIDAFS oscillations, in the energy range above the edge, will
give also direct information on the local composition and on strain
accomodation inside the sticks. Detailed analysis of these oscillations
will be reported elsewhere. 

Finite Difference Method simulations were performed to map the strain
produced by InAs QSs embedded in InP and compare its Fourier Transform
to experimental diffraction intensity maps. We assumed a coherent
growth of InAs on InP \textcolor{black}{(supported by the lack of
dislocations seen in TEM micrography)}. We performed Finite Difference
Method calculations \cite{Niquet98}, in a 2D periodic frame, in the
plane determined by the {[}110{]} and {[}001{]} crystal axis, i.e.
the wires have infinite length along $[1\bar{1}0]$ and a periodic
structure along the {[}110{]}. The QSs length is finite and rather
short (50-200nm), meaning 2D calculations slightly overestimate the
strain, the relaxation along {[}110{]} being not permitted. \noun{}We
chose an FDM cell with a size of $\frac{\sqrt{2}a}{4}\times \frac{a}{4}$.
Then, atoms are placed inside the crystallographic cell according
to Cubic Face Centered Zinc-blende structure, taking into account
elastic deformation. Starting from atomic positions, diffraction intensity
is then calculated using the Distorted Wave Born Approximation \cite{Dosch92}.
No structural disorder has been taken into account. The P and In anomalous
scattering factors were obtained from theoretical values, whereas
experimental values for As in bulk InAs were used. 

\begin{figure}
\subfigure[]{\includegraphics[  width=0.45\textwidth]{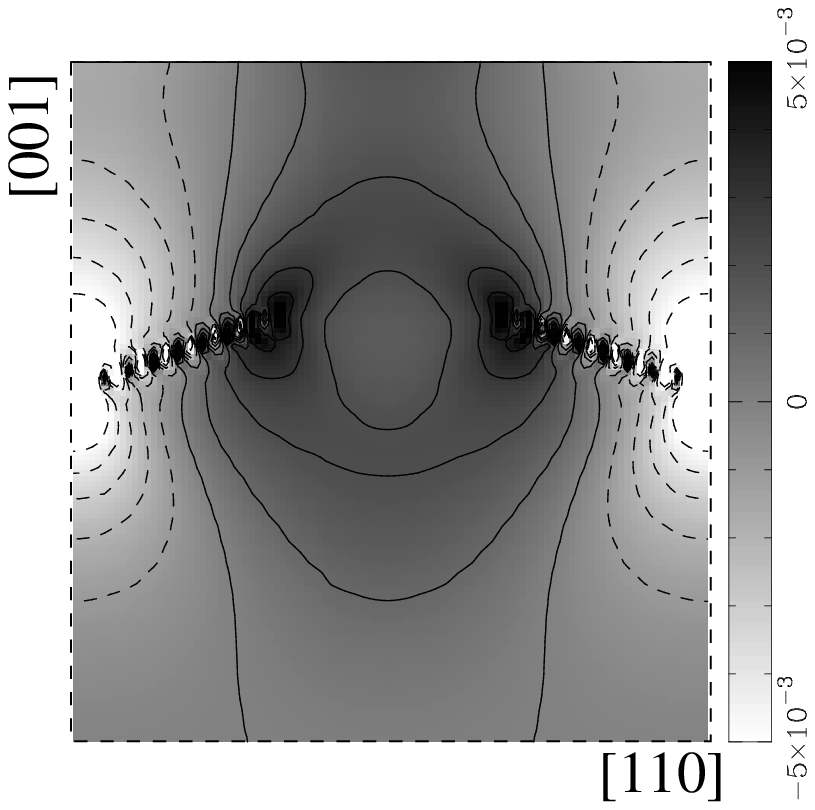}}~~\subfigure[]{\includegraphics[  width=0.45\textwidth]{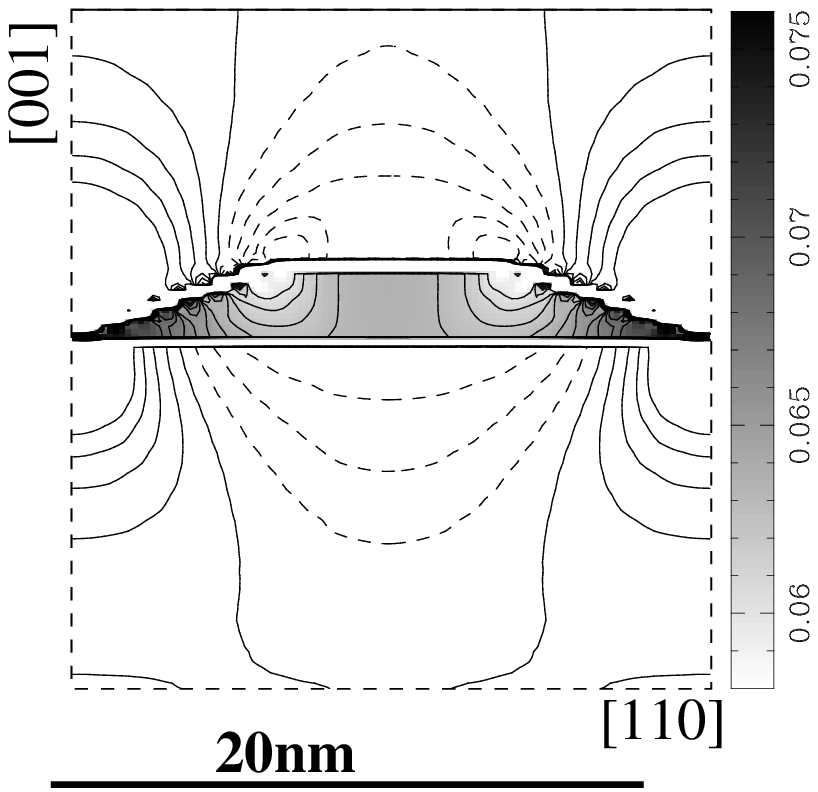}}

\caption{\label{cap:FDM_calculations} FDM simulations of (a) $\varepsilon _{xx}=\frac{a_{ij}-a_{InP}}{a_{InP}}$
and (b) $\varepsilon _{zz}=\frac{c_{ij}-c_{InP}}{c_{InP}}$, where
$a_{ij}$ and $c_{ij}$ are the parameters of cell (i,j). The strain
is relative to bulk InP cell parameter, it is not the relative displacement.
The $\varepsilon _{zz}$ cleary show the morphology of InAs QWrs.
Solid and dashed lines represent positive and negative contour strain,
respectively, starting at $-5.10^{-3}$ and spaced by .001. Ten additional
$\varepsilon _{zz}$ contours are shown inside the wire starting at
0.06. The solid line on top represents the sample surface.}
\end{figure}

Figures \ref{cap:FDM_calculations}a) and \ref{cap:FDM_calculations}b),
show the deformation maps with respect to InP lattice parameter, obtained
with a model of embedded InAs QWrs as deduced from TEM images (fig.
\ref{cap:TEM pictures}b), that well reproduces the diffraction data.
The QWrs side shape is a truncated triangle as shown by TEM. The total
thickness of the QWrs and the capping is equal to $11$nm, as reported
above. Figure \ref{cap:FDM_calculations}b) shows a map of deformations
along the {[}001{]} direction ($\varepsilon _{zz}$), which emphasizes
the morphology of the wires. Indeed the wires are compressed along
{[}110{]} (with partial relaxation shown in fig. \ref{cap:FDM_calculations}a),
thus expanded along {[}001{]}. This explains the sharp contrast of
lattice strain shown in figure \ref{cap:FDM_calculations}b) ($\varepsilon _{zz}$).
Figure \ref{cap:FDM_calculations}a) shows the relative strain $\varepsilon _{xx}$
along {[}110{]}, here it is impossible to disentangle InAs from InP
regions under and on the top of the wires, since the in-plane lattice
strain is continuous at the interface. On figure \ref{cap:FT_FA}a)
are shown the best calculated curves of $F_{A}$ and $F_{T}$, obtained
by optimizing the height and width of the wires, as well as As/P intermixing
at the InAs/InP interface. The splitting of $F_{T}$ is well reproduced
by the simulation and is due to a scattering phase shift between InP
cells above and below the wires. The overall wire thickness and width
are found to be 9 MLs and 22nm, respectively. The simulated ratio
$\beta =\frac{\left\Vert F_{A}\right\Vert }{f_{As}^{0}\left\Vert F_{T}\right\Vert }$
at the maximum of $F_{A}$ is equal to 0.113, i.e close to the experimental
value of 0.124 deduced from the GI-DAFS spectrum (figure \ref{cap:FT_FA}b).
Note that FDM simulation well reproduces the relative positions of
$F_{A}$ and $F_{T}$, with a strain $\varepsilon _{zz}$ of about
6.3\% in the inner part of the wire. For comparison, the strain of
a pseudomorphic InAs thin film grown on InP, as foreseen by the elastic
theory, is 6.7\%. The $\varepsilon _{zz}=\frac{c-c_{InP}}{c_{InP}}=\frac{l_{InP}-l_{F_{A}}}{l_{F_{A}}}$
value, deduced directly from the reciprocal lattice position $l_{F_{A}}$
of the maximum of $F_{A}$ (fig. \ref{cap:FT_FA}a) is equal to 6.1$\pm 0.25$\%,
i.e., within the uncertainty, equal to the value simulated with FDM.
Regarding composition, our FDM simulations show that the QSs inner
part is pure InAs. The experimental curves ($F_{A}$ and $F_{T}$)
are compatible with a weak As/P intermixing at the InP interface,
that would spread over one ML. Indeed, improving the signal-to-noise
ratio together with a 2D data treatment, would allow to fully exploit
the technique sensitivity to map the composition at the ML scale.

In conclusion, we have shown that anomalous diffraction can be used
to extract the structure factor of small size \emph{}InAs nanostructures
\emph{embedded} in InP matrix. This study is a first step towards
a 2D and 3D analysis of anomalous diffraction maps and Grazing Incidence
DAFS data to recover the strain, size, shape and composition of \emph{embedded}
nanostructures.

We are very grateful to F. Né, J.S Micha, S. Arnaud, B. Caillot and
J.F. Bérar for help during the experiments at the beamlines BM32 and
BM2 at the ESRF. M.G.P and H.R. Acknowledge the support of Egide and
Aciones Integradas Programmes (grant ref. HF2002-78) of the French
and Spanish ministries of Research and Education.

\bibliography{/home/renevier/litterature/BiblioTex/dafs,/home/renevier/litterature/BiblioTex/nano}

\end{document}